\documentclass[aps,pra,superscriptaddress,reprint,twocolumn,floatfix]{revtex4-2}
\usepackage[utf8]{inputenc}
\usepackage[T1]{fontenc}
\usepackage[english]{babel}
\usepackage{graphicx}
\usepackage{epsfig}
\usepackage{amsmath,amsfonts,amssymb}
\usepackage{color}
\usepackage{array}
\usepackage[loose]{units}
\usepackage{braket}
\usepackage{xr-hyper}
\usepackage{sidecap}
\usepackage[caption=false]{subfig}
\usepackage{bbold}
\usepackage{physics}
\usepackage{soul}
\usepackage{upgreek}
\usepackage[colorlinks=true,citecolor=blue,linkcolor=magenta]{hyperref}
\usepackage{cleveref}

\graphicspath{{./Figures/}}

\newcommand{\neswarrow}{\mathrel{\text{$\nearrow$\llap{$\swarrow$}}}}
\newcommand{\nwsearrow}{\mathrel{\text{$\nwarrow$\llap{$\searrow$}}}}

\newcommand{\MPINAT}{Max Planck Institute for Multidisciplinary Sciences, D-37077 G\"{o}ttingen, Germany}
\newcommand{\GOE}{University of G\"{o}ttingen, 4th Physical Institute, D-37077 G\"{o}ttingen, Germany}

\begin{document}


\title{Observation of quantum entanglement between free electrons and photons}

\author{Jan-Wilke Henke}
\affiliation{\MPINAT}
\affiliation{\GOE}

\author{Hao Jeng}
\email[]{hao.jeng@mpinat.mpg.de}
\affiliation{\MPINAT}
\affiliation{\GOE}

\author{Murat Sivis}
\affiliation{\MPINAT}
\affiliation{\GOE}

\author{Claus Ropers}
\email[]{claus.ropers@mpinat.mpg.de}
\affiliation{\MPINAT}
\affiliation{\GOE}

\date{\today}


\begin{abstract}
Quantum entanglement is central to both the foundations of quantum mechanics \cite{einstein_EPR_1935,bell_onEPR_1964} and the development of new technologies in information processing, communication, and sensing \cite{nielsen_quantum_2010,giovannetti_advances_2011}.
Entanglement has been realised in a variety of physical systems, spanning atoms, ions, photons, collective excitations, and hybrid combinations of particles \cite{freedman_Experimental_1972,kwiat_SPDC_1995,hagley_Generation_1997,turchette_Deterministic_1998,blinov_Observation_2004,togan_Quantum_2010,pashkin_Quantum_2003,julsgaard_Experimental_2001,delteil_Generation_2016,lee_Entangling_2011,lin_Quantum_2020,holland_On-demand_2023,bao_Dipolar_2023,atlas_Observation_2024}.
Remarkably, however, photons and free electrons---the quanta of light and their most elementary sources---have never been observed in an entangled state.
Here, we demonstrate quantum entanglement between free electrons and photons.
We show that entanglement is produced when an electron, prepared in a superposition of two beams, passes a nanostructure and generates transition radiation in a polarisation state tied to the electron path.
By implementing quantum state tomography, we reconstruct the full density matrix of the electron-photon pair, and show that the Peres-Horodecki separability criterion is violated by more than 7 standard deviations. 
Based on this foundational element of emerging free-electron quantum optics, we anticipate manifold developments in enhanced electron imaging and spectroscopy beyond the standard quantum limit \cite{henke_Probing_2025, kazakevich_Spatial_2024,rembold_StateAgnostic_2025a,dwyer_Quantum_2023,kruit_Designs_2016,koppell_Transmission_2022}.
More broadly, the ability to generate and measure entanglement opens electron microscopy to previously inaccessible quantum observables and correlations in solids and nanostructures.
\end{abstract}

\maketitle


\section*{Introduction}
Since the advent of quantum mechanics one hundred years ago, the interaction between a free electron and the electromagnetic field has served as a paradigm of fundamental quantum phenomena.
Beyond guiding early developments in quantum theory, this basic system also became the foundation for quantum electrodynamics, described by Feynman as the ``jewel of physics'', which has played an instrumental role in our understanding of the non-classical behaviour of nature.
A range of seminal experiments has illustrated quantum characteristics of electron-photon interactions.
Notable examples include Compton scattering,  which provided early evidence for the particle-like behaviour of light \cite{compton_Quantum_1923}, and precision measurements of the anomalous magnetic dipole moment, presenting some of the most stringent tests of quantum electrodynamics to date \cite{fan_Measurement_2023}. 

In more recent developments, free electrons were made to emit and absorb integer numbers of photons when interacting with laser fields \cite{barwick_photon-induced_2009}, with inelastic scattering mediated by the dielectric or plasmonic response of nanostructures \cite{park_Photoninduced_2010,garcia_de_abajo_multiphoton_2010,piazza_simultaneous_2015,henke_integrated_2021,dahan_imprinting_2021,gaida_Attosecond_2024}.
Such stimulated interactions can drive multilevel Rabi oscillations \cite{feist_quantum_2015}, and facilitate the precise control of the quantum states of electrons in space and time \cite{freimund_Bragg_2002,priebe_Attosecond_2017a,morimoto_Diffraction_2018,vanacore_Attosecond_2018,madan_Ultrafast_2022,gaida_Attosecond_2024,fang_Structured_2024}.
Spontaneous inelastic scattering in the absence of a laser, on the other hand, produces correlations between electron beams and the radiation it generates \cite{bendana_single-photon_2011,feist_cavity-mediated_2022,varkentina_cathodoluminescence_2022,arend_Electrons_2024}.
The combination of both stimulated and spontaneous processes is predicted to allow for the generation of nearly arbitrary quantum states of light \cite{ben_hayun_shaping_2021,dahan_creation_2023,zhang_Spontaneous_2024,di_giulio_Tunable_2025}.

Whilst the quantum-mechanical nature of the interactions is evident in the examples above, it is an entirely different matter to prepare and observe free electrons entangled with the electromagnetic field. Quantum entanglement manifests as correlations between the quantum states of subsystems, but only when such correlations are measured directly can entanglement be verified.
Bound electrons in atoms and solid-state structures have long provided avenues for quantum information research.
Free electrons, in contrast, form the basis of electron microscopy and benefit from decades of technological development, offering exceptional beam control, high coherence, and advanced methods of detecting their interactions with electromagnetic excitations \cite{garcia_de_abajo_optical_2010, polman_electron-beam_2019}.
Nonetheless, entanglement involving free electrons has thus far remained a subject of theoretical investigations and proposals.

The quantum eraser is one of the most iconic thought experiments that illustrates the defining features of entanglement \cite{ma_Delayedchoice_2016}.
In the quantum eraser scheme originally proposed by Scully and Drühl \cite{scully_Quantum_1982}, interference between photons emitted by two separated atoms depends on whether which-path information is available.
If the atoms end in indistinguishable states, interference is observed; if their final states are distinguishable, the interference disappears.
However, by detecting an additional photon in a way that ``erases'' this which-path information, interference can be restored in the correlations between the photons.

In this study, we realise quantum entanglement between a free electron and a single photon, in a configuration inspired by the quantum eraser, employing electron path and photon polarisation degrees of freedom.
Electrons are prepared in a superposition of two beams passing a nanostructure, generating photons that mark the position of the beams, with the marker photons distinguished by their polarisation states.
If we assume for the moment that these polarisations are the diagonal states $\ket{\neswarrow}$ and $\ket{\nwsearrow}$, the electron becomes entangled with the photon, and their joint quantum state is given by the Bell state,
\begin{equation}
    \ket{\psi} = \frac{1}{\sqrt{2}}(\ket{L, \neswarrow} + \ket{R, \nwsearrow})~.
\end{equation}
Suitable polarisation measurements could either keep the information held by the marker photons, or erase it by probing different superpositions of the polarisation states.
By observing how electron interference fringes respond to coincident measurements of the photons in different polarisation bases, we can reconstruct the density matrix for the electron-photon pair and perform a rigorous test for the presence of entanglement.


\section*{Experiment}
\label{sec:scheme}

\begin{figure}[thb]
    \centering
    \includegraphics[width=\columnwidth]{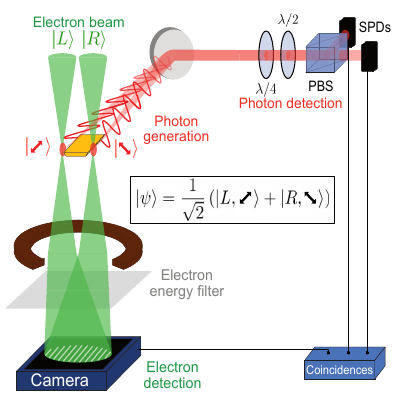}
    \caption{\textbf{Quantum eraser-type scheme for entangling free electrons and photons.}
        A coherently split electron beam (green) with electrons in a superposition of the left and right beam, denoted $\ket{L}$ and $\ket{R}$, respectively, passes a sample structure (yellow).
        Upon traversing the structure, electrons can generate a photon (red) whose polarisation, illustrated by small red arrows, depends on the side of the sample the electron passes by.
        Under ideal conditions, this results in the formation of an entangled Bell state for the joint electron-photon state, expressed in the inset.
        To verify electron-photon entanglement, the photons are collected and guided to an optical setup consisting of a quarter-wave plate ($\lambda/4$), a half-wave plate ($\lambda/2$), a polarising beam splitter (PBS) and two single-photon detectors (SPDs), enabling polarisation control and state-sensitive detection.
        Electrons are detected by a hybrid-pixel camera (dark blue) behind a projection lens and an electron energy filter (grey) transmitting electrons that lost energy by the generation of a photon. Interference patterns formed by electrons detected in coincidence with photons yield the correlations necessary to prove entanglement.
    }
    \label{fig:1}
\end{figure}

Experiments were carried out with metal-coated glass prisms placed in a transmission electron microscope (TEM) as illustrated in Figure~\ref{fig:1}.
The microscope is equipped with a cold field-emission source, and the electrons are divided into multiple beams using an amplitude grating (\cite{harvey_efficient_2014,johnson_scanning_2021}; see Methods and Supplementary Information), with an additional aperture placed below the sample plane to block all beams but two.
The sample is about \unit[500]{nm} in width, $\unit[2]{\mu m}$ in length, and is carved out of the tip of a single-mode optical fibre using focused ion-beam milling.
Coherent cathodoluminescence \cite{garcia_de_abajo_optical_2010,polman_electron-beam_2019,taleb_Phaselocked_2023}, more specifically transition radiation, is collected by the optical fibre and extracted from the microscope column for analysis: A quarter-wave plate and a half-wave plate are placed at the output of the optical fibre, followed by a polarising beamsplitter and a pair of superconducting nanowire single-photon detectors.
The overall generation and collection efficiencies are estimated to be about $10^{-6}$ photon counts per electron.
Coincidence detection between electrons and the generated photons is possible through a pixelated event-based electron detector as discussed in detail in Refs. \cite{feist_cavity-mediated_2022,varkentina_cathodoluminescence_2022,arend_Electrons_2024}. To avoid detector saturation, electrons are energy-filtered to the range of \unit[1.5-2.0]{eV} energy loss, largely covering to the spectral range of the generated photons.

\begin{figure*}[thb]
    \centering
    \includegraphics[width=\textwidth]{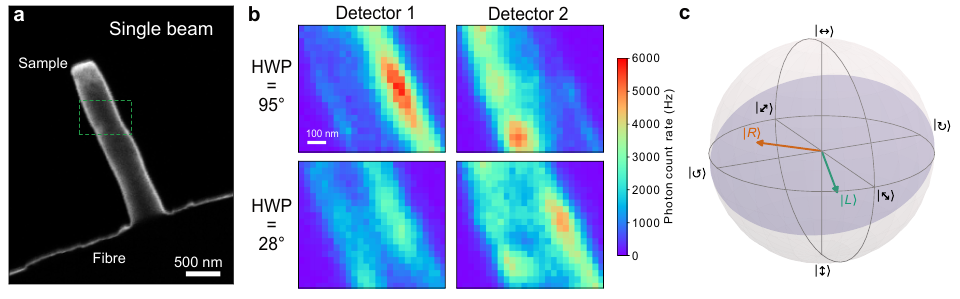}
    \caption{\textbf{Characterisation of polarised cathodoluminescence.}
        \textbf{a} Dark-field image of the sample structure, cut from the core of a single-mode fibre and coated with a thin ($\approx \unit[40]{nm}$) metal layer.
        A single electron probe is raster-scanned over the part of the sample marked by the green rectangle.
        \textbf{b} Maps of the photon count rates in the output ports of the polarising beam splitter when scanning a single electron probe over the sample structure.
        For a half-wave plate setting of $95^{\circ}$ (QWP$=30^{\circ}$), the electron beam passing on the right primarily generates photons registered on detector 1.
        In contrast, the beam on the left generates photons that are primarily detected by detector 2.
        At a half-wave plate angle of $28^{\circ}$ (QWP$=30^{\circ}$), however, the detection of a photon on each of the detectors is equally likely for a beam passing the structure on the left or the right.
        \textbf{c} Electrons passing on different sides of the structure, as employed in the interference measurements, generate photons of different polarisation.
        The corresponding polarisation states can be visualised on the Bloch sphere with the colour and label denoting the electron beam position.
    }
    \label{fig:2}
\end{figure*}

We operate the electron microscope in two complementary modes: a scanning mode where a single electron beam is raster-scanned over the sample to map the position-dependent photon generation rate and polarisation, and a dual-beam mode where we hold the two beams at a fixed position near the sample and observe their far-field interference.
The scanning region for the single-beam measurements is highlighted by the green box in Figure \ref{fig:2}a, which shows the dark-field image of the sample.
We observe polarised cathodoluminescence, with distinct polarisation states on opposite sides of the sample (cf. Fig. \ref{fig:2}b-c).
This is evident in the spatial maps in Figure \ref{fig:2}b, which show count rates at both detectors for two different settings of the wave plates.
Specifically, the wave plate setting for the top row (HWP=$95^\circ$) clearly divides the photons from both sample edges into the detectors 1 and 2, while the bottom row shows similar count rates at both edges and for both detectors.
Performing such raster-scans for a comprehensive set of quarter-wave plate and half-wave plate angles between $\unit[0^{\circ}-90^{\circ}]{}$, the electron-position dependent photonic quantum state is determined through maximum likelihood estimation \cite{lvovsky_Iterative_2004}.
Figure \ref{fig:2}c displays a Bloch-sphere representation of the photon states retrieved for the two beam positions used in the following interference measurements.
The principal axis of their states aligns approximately with the diagonal polarisation basis, with an angle of about $121^\circ$ between the two Bloch vectors, yielding reasonable distinguishability.

\begin{figure*}[thb]
   \centering
   \includegraphics[width=\textwidth]{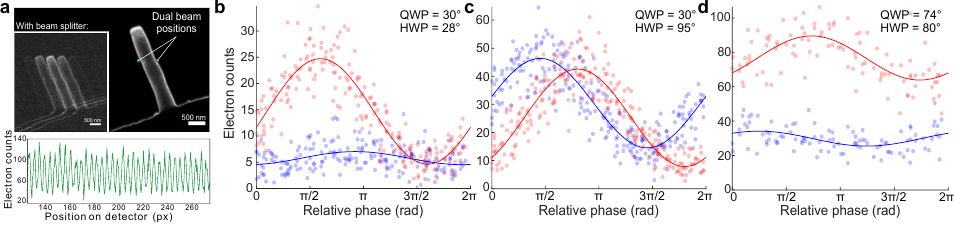}
   \caption{\textbf{Electron interference fringes conditioned on photon polarisation measurement.}
        \textbf{a} With the electron beam splitter installed, the superposition of electron beams leads to replicas of the sample in the dark-field image (inset).
        Two of the beams are positioned on either side of the sample (green circles).
        Recombining these dual beams on the detector behind the energy filter and selecting the coincidences with one of the detectors yields the interference pattern shown at the bottom.
        \textbf{b-d} Coincidence-gated interference patterns for three sets of polarisation measurements with quarter-wave plate and half-wave plate angles of $(30^\circ,28^\circ)$ (b), $(30^\circ,95^\circ)$ (c), and $(74^\circ,80^\circ)$ (d), respectively, corresponding approximately to measurements along three mutually orthogonal axes on the Bloch sphere. Pairs of interference fringes heralded by each detection outcome in different polarisation bases are distinguished by the colours red and blue. The horizontal axis indicates the relative phase between the two electron beams at the location of the electron detector, and the vertical axis gives the recorded electron counts. The fitting of sinusoidal functions to the observations is shown as solid lines.}
   \label{fig:3}
\end{figure*}

In the interference measurements, we prepare the electrons in a two-beam superposition, with a distance adjusted such that each beam passes through a region of high photon generation rate at the sample edges (green dots in Fig.~\ref{fig:3}a).
Proper beam separation is confirmed by inspecting the appearing copies of the sample in scanning dark-field images (cf. inset in Fig. \ref{fig:3}a).
Recombining the far field of both beams on a detector behind the energy filter, we record interference patterns of electrons coincident with photons arriving at either of the detectors (see Methods and SI).
An example interference pattern is plotted at the bottom of Fig. \ref{fig:3}a.

To carry out quantum state tomography of the correlated electron-photon pair, at least three sets of photon polarisation measurements are necessary; the quarter-wave plate and half-wave plate angles were thus set to $(30^\circ,28^\circ)$, $(30^\circ,95^\circ)$, and $(74^\circ,80^\circ)$, which correspond approximately to three mutually orthogonal axes on the Bloch sphere (see Fig. \ref{fig:3}b-d).
It is found that the conditional electron interference fringes agree well with sinusoidal fits, and display variations in visibility between about $14.5\%$ to $68.7\%$.
Relative phase shifts vary between $16.2\%$ and $24.5\%$ of a cycle, depending on the detected photon polarisation. These interference fringes demonstrate coherent cathodoluminescence from a transversely delocalised electron probe.
Moreover, the observed phase shifts and distinct visibilities represent a heralding of different electron quantum states.
Importantly, these conditional interference fringes, combined with the cathodoluminescence data above, are sufficient for obtaining a tomographic reconstruction of the joint density matrix of the electron-photon state.
In order to make use of all of the data in the reconstruction, as well as to directly facilitate error estimation, we opt for a Bayesian approach (see Methods). 

\begin{figure*}[thb]
   \centering
   \includegraphics[width=0.82\textwidth]{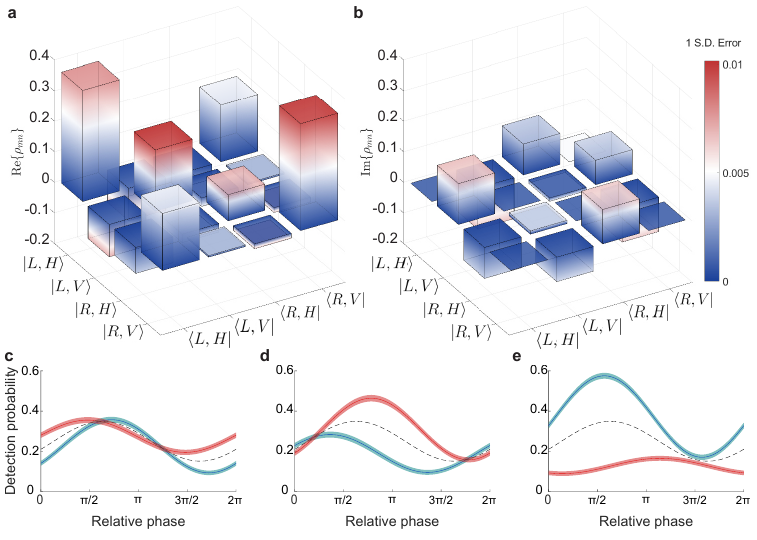}
   \caption{\textbf{The electron-photon joint density matrix reconstructed using Bayesian estimation.}
       \textbf{a,b} Real (\textbf{a}) and imaginary part (\textbf{b}) of the reconstructed electron-photon density matrix. The colourmaps indicate 1 standard deviation (SD) error, with a scale shown on the right. For this visualisation, the photon basis was rotated to optimise the fidelity relative to the Bell state $(\ket{L,H}+\ket{R,V})/\sqrt{2}$.
       \textbf{c-e} Inferred electron interference fringes when photon Z, X, and Y measurements are carried out in coincidence in this rotated basis. The mean values of the fringes are shown in dashed lines for comparison, while the shaded areas denote the 3 SD error.
       }
   \label{fig:4}
\end{figure*}

We have observed a mean posterior density matrix for the electron-photon pair that bears close similarity with a Bell state (see Fig. \ref{fig:4}a\&b).
To represent these density matrices in an intuitive form, a local unitary operation has been applied to the photonic basis such that the highest fidelity with respect to the Bell state $(\ket{LH}+\ket{RV})/\sqrt{2}$ is obtained, and its value is determined as $F = 0.543 (0.008)$.
In this basis, the main polarisation axis that distinguishes the photons from the two beams is aligned closely to the Z axis of the Bloch sphere.

As an alternative depiction of the quantum correlations in the joint electron-photon state, we display the behaviour reminiscent of a quantum eraser in this basis (Figs.~\ref{fig:4}c-e).
Because the axes of the photon polarisation measurements are tilted relative to the axes in which the Bell correlations are most apparent, the raw interference fringes are not optimal for visualising the basis-dependent separation of interference fringes.
However, this information is readily obtained from the posterior distribution of the density matrices.
The interference fringes in the optimised basis shown in Fig. \ref{fig:4}c-e show quite clearly that different measurements provide different levels of which-path information.
In the z-basis (Fig.~\ref{fig:4}c), relatively minor visibility and detector differentiation for both electron fringes are found, while the interference is enhanced if the photons are measured in bases that do not carry this information (cf. Fig. \ref{fig:4}d\&e).
The degree of correlations between these interference fringes and the photon polarisation is directly linked to the fidelity $F$ with respect to the Bell state.

\begin{figure*}[thb]
   \centering
   \includegraphics[width=0.6\textwidth]{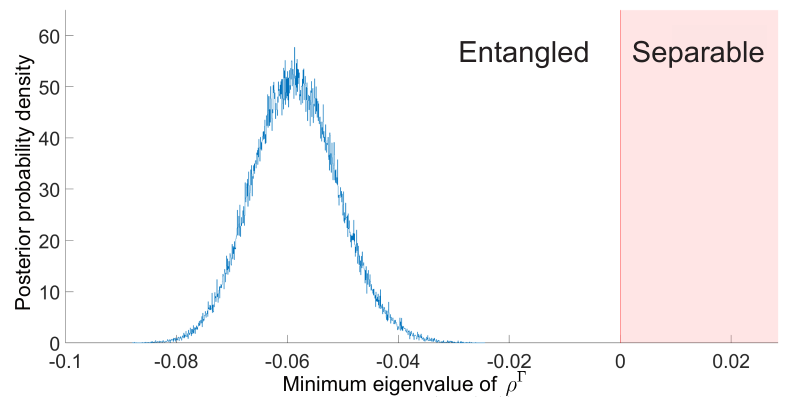}
   \caption{\textbf{Entanglement test using the Peres-Horodecki criterion.} The blue trace shows the posterior distribution of the minimum eigenvalues of the partially transposed density matrix. The Peres-Horodecki separability criterion for two-qubit systems states that this quantity is non-negative if and only if the system is separable.
   }
   \label{fig:5}
\end{figure*}

The mean fidelity with Bell states exceeding $0.5$ by 5 standard deviations directly implies the presence of quantum entanglement.
However, such fidelity-based entanglement witnesses are not optimal for quantum states somewhat further from Bell states, and can underestimate the statistical significance of the results.
Therefore, we have also carried out a test for quantum entanglement using the Peres-Horodecki separability criterion: the minimal eigenvalue of the partially transposed density matrix was determined to be $-0.059(0.008)$, and represents a violation of the criterion by more than 7 standard deviations as shown in Figure \ref{fig:5}.
For two-qubit states, this quantity is directly linked to other entanglement measures such as the concurrence, $0.14(0.02)$, and the entanglement of formation, $0.05(0.01)$.

In the present experiments, we identify three sources of decoherence that limit the achieved degree of entanglement: the finite coherence of the initial electron state, the distinguishability and coherence of photons emitted by the two beams, and the decoherence of the electron beam as it crosses the nanostructure.
All but the first of these factors are tied intrinsically to the nature of the electron-photon interaction and the structure at hand, and are difficult to address independently from other degrees of freedom.
However, the coherence of the initial electron beam is well characterised and can be corrected for.
Full mathematical details of an optional correction procedure are described in the Methods.
Briefly, if we assume that the transition radiation occurs independently for each electron beam, then we can infer the value of any expectation value of quantum observables for an ideal, perfectly coherent initial electron state, provided that the corresponding expectation value was measured in the experiment with the real, imperfect electron probe.
Using the above procedure, and with an initial electron coherence of $0.73$ extracted from electron interference fringes in the absence of the nanostructure (see Supplementary Information), we deduce a coherence-corrected value of the Bell state fidelity of $F = 0.64 (0.01)$.
Such improved states will therefore be available experimentally.

\section*{Discussion \& Conclusion}
The observations above paint the picture of a quintessential quantum eraser operating with electrons.
The interaction of the electrons with the nanostructure acts as a quantum measurement of the position of the electron: in the absence of the nanostructure, the two beams will interfere and give rise to clear interference patterns, but in the presence of the nanostructure, the electrons will emit, with some probability, marker photons that have polarisations entangled with the electron path. The subsequent observation of electron interference fringes is then conditional on the basis in which we probe the photon markers. The simultaneous measurement of the quantum-mechanical electron and photon states yields a direct observation of their entanglement.

Based on our findings, a free electron coupled to a single photon can now be isolated, manipulated, and measured together as a single hybrid quantum system.
As indicated by numerous theoretical proposals \cite{kfir_entanglements_2019,konecna_entangling_2022,baranes_free_2022,haindl_Coulomb_2023,kazakevich_Spatial_2024,henke_Probing_2025,rembold_StateAgnostic_2025a}, it is natural to expect that many different kinds of entangled states could be produced through variants of our technique, by carrying out measurements in different photonic degrees of freedoms such as momentum, angular momentum, and various spatial and spectral modes.
Furthermore, it is also straightforward to scale the system up to multipartite entangled states and higher-dimensional entangled states by increasing the number of electron beams generated by the amplitude grating.
Notably, this is possible because the mutual coherence of the discrete, well-separated probes produced by the grating is not reliant on a high degree of global beam coherence in each of the individual probes.
This also differentiates the present approach from inelastic holography using biprisms~\cite{lichte_Inelastic_2000}, an approach that has previously stimulated the consideration of entanglement in electron microscopy \cite{potapov_Inelastic_2007,mechel_Quantum_2021}.

Owing to the central role that quantum entanglement plays in quantum information and technology, observation of entanglement leads to numerous possibilities for integrating quantum applications into electron microscopy.
A well-known example is the reduction of radiation damage below the standard quantum limit using the methods of quantum metrology \cite{kruit_Designs_2016, koppell_Transmission_2022}.
Whilst the realisation of such a quantum electron microscope is believed to still be some time away, the essential ingredients are now within reach.
Path-entangled states of two or more electrons \cite{haindl_Coulomb_2023}, for instance, may allow for quantum-enhanced phase contrast measurements, and these can now be produced through entanglement swapping of electron-photon entangled pairs \cite{henke_Probing_2025}.
Finally, by generalising the quantum state tomography procedure implemented here to quantum process tomography through the use of tailored electron probes, it will become possible not only to produce entanglement but also to characterise the fundamental entangling interactions that take place within a nanostructure, thus offering a valuable tool for the direct study of quantum behavior of materials at the nanoscale.

\section*{Methods}
\label{sec:methods}

\begin{footnotesize}
\subsection*{TEM Setup}
The experiments are performed using a JEOL JEM-F200 TEM fitted with a cold-field emission electron source, yielding a continuous electron beam with an initial energy spread of about \unit[500]{meV} at a centre electron energy of \unit[80]{keV}.
A detailed drawing of the experimental setup is provided in the supplemental Figure \ref{fig:SI1}.
The microscope is equipped with a custom condenser aperture holder allowing for the placement of standard apertures of \unit[40]{$\mu$m} and \unit[100]{$\mu$m} diameter as well as a \unit[40]{$\mu$m} x \unit[40]{$\mu$m} amplitude grating.
The amplitude grating is used to produces multiple electron probe beams, similar to grating-type phase plates employed in inelastic holography and electron interferometry \cite{harvey_efficient_2014,johnson_scanning_2021}, and consists of \unit[130]{nm}-wide slits with a pitch of \unit[300]{nm}.
The grating pattern is milled into a gold-coated (\unit[100]{nm} film thickness)  \unit[50]{nm}-thick silicon nitride membrane (\unit[50]{$\mu$m} x \unit[50]{$\mu$m} window) via focused ion beam etching (cf. supplemental Figure \ref{fig:SI1}b).
Using the low-magnification STEM mode, multiple probe beams of approx. \unit[510]{nm} spacing and $<\unit[20]{nm}$ diameter are formed in the sample plane when the grating is inserted, as quantified by imaging the spots on a camera in front of the electron imaging spectrometer (see supplemental Figure \ref{fig:SI1}d).
The probe spacing can be fine-tuned by adjusting two of the condenser lenses in a telescope setting to match the width of the sample, yielding an intensity ratio of $0.64$ to $0.36$ in the 0th and 1st grating diffraction orders.
In the measurements, the electron beams are kept stationary at positions on the structure optimised for photon generation rates and electron transmission.
In order to block the electron probes that are not interacting with the sample, an objective lens aperture with \unit[1.2]{$\mu$m} diameter is placed below the interaction region (see Figure \ref{fig:SI1}a\&e).
In the performed characterisation measurements, a standard \unit[40]{$\mu$m} condenser aperture is used resulting in a single probe ($<\unit[20]{nm}$ diameter) that is scanned across the sample with a scanning resolution of \unit[25]{nm} and an integration time of \unit[100]{ms} per scan pixel.

A post-column imaging energy filter (CEFID, CEOS) equipped with a hybrid-pixel electron detector (EM CheeTah T3, Amsterdam Scientific Instruments) is employed to analyse the electron beam interacting with the sample.
Electrons entering the spectrometer are energy dispersed and filtered by an energy-selective slit of less than \unit[20]{meV} bi-sided r.m.s. non-isochromaticity (energy window edge sharpness), allowing the selection of electrons that have lost the energy equivalent of the generated photons.
For the eraser measurements, this energy loss window ranges from \unit[1.5-2.0]{eV}, corresponding to photon wavelengths of approximately \unit[620-830]{nm}.
The individual, energy-filtered electrons are then imaged on the detector and registered by their position and arrival time.
Electron-photon correlation measurements are enabled by combining this information with the detection of photons, connecting the output of the single-photon detectors to the time-to-digital converters on the hybrid-pixel detector.

\subsection*{Optical Setup}
Photons generated by the electron beam are collected through a 780HP optical single-mode fibre and coupled to a free-space polarisation control consisting of a broadband (\unit[700-1200]{nm}) quarter-wave and half-wave plate.
A broadband (\unit[650-1000]{nm}) polarising beam splitter is employed to split the photons according to their polarisation state before forwarding them to two superconducting nanowire single-photon detectors (Single Quantum Eos CS) via two fibre polarisation controls.
These single-photon detectors feature a detection efficiency of >\unit[75]{\%} in the wavelength range of \unit[785-850]{nm} with a dark-count rate of <\unit[10]{Hz} and the transmission in the two optical paths is characterised to be \unit[68]{\%} and \unit[76]{\%}, respectively.
Their output signal is connected to the time-to-digital converter of the electron hybrid pixel detector that acts as a time tagger and allows for electron-photon correlation analysis.
To obtain the photon count rate maps, the single electron probe is raster-scanned over a $\unit[25]{px} \times \unit[30]{px}$ area and photon counts are integrated for \unit[100]{ms} per pixel at different QWP and HWP settings of $\unit[0^{\circ}-90^{\circ}]{}$ in steps of $10^{\circ}$.
Electron-photon coincidence data is collected with the two beams placed at the sample edges for at least \unit[440]{s} per setting.

\subsection*{Sample Preparation}
The sample is fabricated from a commercial 780HP single-mode optical fibre tapered to <\unit[10]{$\mu$m} diameter under a \unit[75]{$^\circ$} angle and sputter-coated with >\unit[100]{nm} gold.
The tapered fibre end is polished to give a \unit[30]{$\mu$m} diameter endface, before using focused ion beam (FIB) milling to cut a prism-shaped plateau into the core of the optical fibre.
This prism structure features two faces angled \unit[45]{$^\circ$} relative to the incident electrons and a base length of \unit[450-550]{nm}.
After processing with the FIB, the structure and tapered part of the fibre are coated with approx. \unit[40]{nm} of aluminium to prevent charging-related deflection of the electron beam in the TEM and enhance the generation of transition radiation from the prism faces.
A fraction of the photons generated by electron-driven coherent cathodoluminescence is coupled to the guided mode of the optical fibre. 
The generation probability of \unit[$1.5 \cdot 10^{-6}$]{} for such a guided photon can be estimated from the photon and electron count rates.
The emission spectrum of the electron-driven photon generation was characterised at an electron energy of \unit[200]{keV} using a grating spectrometer (see spectrum in supplemental Figure \ref{fig:SI2}), showing the same bandwidth and characteristics on both sides of the sample structure.

\subsection*{Data Processing}
In a first step, hits registered on the Timepix detector are grouped in electron clusters with a time of arrival as well as a position on the detector.
The resulting electron clusters are then correlated in their arrival time to the detection of photons on the two detectors and electrons within a coincidence time window of about \unit[5]{ns}, featuring a peak with a signal-to-noise ratio >\unit[4]{}, are selected to yield the coincidence interference patterns (cf. supplemental Figure \ref{fig:SI3}a).
These patterns are obtained by binning the coincidencce events in $\unit[1028]{px} \times \unit[1028]{px}$ spatial bins, correcting for the uncorrelated background which is comprised of unscattered electrons in the zero-loss peaks tail as well as loss-electrons whose corresponding photon was not detected, and rotation to align the fringes (cf. supplemental Figure \ref{fig:SI3}b-e).
The finally extracted (by summation and sorting by phase) line plots are shown in Figure \ref{fig:3}.

\subsection*{Bayesian Quantum State Tomography}
Each measurement outcome of the photon polarisation and the electron position or relative phase can be assigned a projection operator $\xi_j$ acting on the joint density matrix.
We can thus form the likelihood function for the dataset $L(\rho) = tr(\xi_j\rho)^{N_j}$, where $N_j$ is the observed number of counts for that measurement.
The posterior distribution of the density matrix given these observations is then expressed by:
\begin{equation*}
    \pi(\rho) = L(\rho)\pi_0(\rho),
\end{equation*}
where $\pi_0$ is the prior and will be taken to be a uniform distribution.
In general, it is difficult to obtain Bayesian estimates directly from this expression owing to the complicated structure of the space of density matrices.
Instead, we parameterise this space using complex vectors as described in Ref. \cite{lu_Bayesian_2022}, with all the requirements of physicality of the state automatically built into this parametrisation.

Bayesian reconstruction of density matrices is carried out by drawing samples from the posterior distribution using Markov chain Monte Carlo \cite{kim_Bayesian_2023}.
Specifically, we use the Metropolis-Hastings update rule: given the current state of the chain $m_i$ and a proposal function $q(\cdot|\cdot)$, we first draw a random sample $m'$ from the distribution $q(\cdot | m_i)$ and accept this proposed sample as the next state in the chain $m_{i+1}$ with probability 
\begin{equation*}
    \alpha = \min\left(1,\frac{\pi(m')q(m_i|m')}{\pi(m_i)q(m'|m_i)}\right)~.
\end{equation*}
If the proposed sample is rejected, we set $m_{i+1} = m_i$.
For the proposal function, we use preconditioned Crank Nicholson:
\begin{equation*}
    q(m|m_i) = \mathcal{N}(\sqrt{1-\beta^2}m_i, \beta^2 \mathbb{1}),
\end{equation*}
where $\mathcal{N}$ denotes the normal distribution and $\beta$ is an adjustable parameter.
In practice, $\beta$ is always chosen such that the acceptance rate of the Markov chain Monte Carlo is close to the optimal value of about $23.4\%$. 

In order for samples drawn from the posterior distribution to represent it accurately, the Markov chains must have converged.
Extended data Figure SI4 in the Supplementary Information illustrates this convergence.
Furthermore, the samples drawn should show a minimal amount of autocorrelation in order for the results to be statistically significant; this is also shown in the extended data Figure \ref{fig:SI5}d.
Here, we found that a chain of about $10^6$ samples is more than sufficient to obtain convergence and at the same time provide a large number of independent samples.
A burn-in of $10\%$ is applied to the chains.

We note that the reconstruction seems to slightly overestimate the interference visibility in the measurement with wave plate settings (QWP$=\unit[74^{\circ}]{}$, HWP$=\unit[80^{\circ}]{}$) (cf. supplemental Figure \ref{fig:SI5}e), while the visibilities in the other two measurements are slightly undererstimated (cf. supplemental Figure \ref{fig:SI5}c\&d).
We attribute this to a reduced fringe visibility in the (QWP$=\unit[74^{\circ}]{}$, HWP$=\unit[80^{\circ}]{}$) measurement most likely caused by carbon deposition on the sample structure throughout the electron beam illumination, as the mentioned data was collected at a later time.

\subsection*{Coherence Correction}
Suppose that the generation of photons by the incident electron satisfies the following properties:
\begin{itemize}
    \item It can be described as a completely-positive trace-preserving map,
    $\rho \rightarrow \mathcal{E}(\rho)$,
    where $\rho$ and $\mathcal{E}(\rho)$ are the density matrices of the incident electron and the joint electron-photon states, respectively.
    \item The action of $\mathcal{E}$ on the states $\ket{0}$ and $\ket{1}$ are known:
     \begin{align*}
        \mathcal{E}(\ket{0}\bra{0}) &= \ket{0}\bra{0} \otimes \rho_0\\
        \mathcal{E}(\ket{1}\bra{1}) &= \ket{1}\bra{1} \otimes \rho_1
    \end{align*}
    with $\rho_0$ and $\rho_1$ describing local photon states.
\end{itemize}
Then the expectation value $tr(\mathcal{O}\mathcal{E}\rho_{in})$ measured in the experiment for an imperfect electron beam $\rho_{in}$ could be inverted to obtain the expectation value $tr(\mathcal{O}\mathcal{E}\rho)$ for a perfectly coherence electron beam $\rho$. 

Specifically, if we write out explicitly the density matrices,
\begin{equation*}
  \rho_{in} = \begin{bmatrix}
  a & c \\
  \overline{c} & b 
  \end{bmatrix}~,
\end{equation*}
and
\begin{equation*}
  \rho = \begin{bmatrix}
  a' & c' \\
  \overline{c'} & b'
  \end{bmatrix}~,
\end{equation*}
then the two expectation values are related by the equation:
\begin{align*}
    tr(\mathcal{O}\mathcal{E}\rho) &= (tr(\mathcal{O}\mathcal{E}\rho_{in}) - a \Tr(\mathcal{O}\mathcal{E}(\ket{0}\bra{0})) - b \Tr(\mathcal{O}\mathcal{E}(\ket{1}\bra{1})))\frac{c'}{c} \\
    &+ a' \Tr(\mathcal{O}\mathcal{E}(\ket{0}\bra{0})) + b' \Tr(\mathcal{O}\mathcal{E}(\ket{1}\bra{1})).
\end{align*}

For the coherence-corrected expectation values reported in this work, we put $a' = a$, $b' = b$, and $c' = \sqrt{ab}$, corresponding to an electron state with the same populations in either beam but perfect spatial coherence.
The impact of the coherence correction on the Bell state fidelity is shown in Figure \ref{fig:SI6}.

\end{footnotesize}


\subsection*{Supplementary information}
The Supplementary Information contains Sections 1–3, including Supplementary Figures 1–6. Section 1: Experimental setup; Section 2: Sample characterisation; Section 3: Data processing \& evaluation.

\subsection*{Author Contributions}
J.W.H. and M.S. designed and fabricated the fibre optical samples.
J.W.H. built the setup and performed the experiments.
H.J. and J.W.H analysed the data..
The study was planned and directed by C.R..
The manuscript was written by H.J., J.W.H and C.R., after discussion with and input from all authors.

\subsection*{Acknowledgements}
It is our pleasure to acknowledge Jörg Malindretos for metal coating of the samples, Nora Bach and Tim Dauwe for fabrication of the amplitude grating, and Tobias Meyer for providing laboratory space for the single photon detectors.
We also wish to thank Valerio Di Giulio, Hugo Louren\c{c}o-Martins, Till Domröse, F. Javier Garc\'ia de Abajo, Ofer Kfir, Armin Feist, Germaine Arend and Tyler Harvey for insightful discussions.
During preparation of this manuscript, we became aware of an independent research project on related aspects of electron-photon entanglement, in the group led by Philipp Haslinger.

\subsection*{Funding Information}
This work was funded by the Deutsche Forschungsgemeinschaft (DFG grant number 432680300/SFB 1456, project C01), the Gottfried Wilhelm Leibniz Programme, and Horizon 2020 Research and Innovation Programme of the European Union (grant agreement number 101017720: FET-Proactive EBEAM).

\subsection*{Data Availability Statement}
The data used to produce the figures in this work will be made available in an online repository upon publication of this work.

\subsection*{Competing Interests}
The authors declare no competing financial interests.


\bibliographystyle{unsrt}
\bibliography{QE_experiment_arxiv}

\begin{thebibliography}{10}

\bibitem{einstein_EPR_1935}
A.~Einstein, B.~Podolsky, and N.~Rosen.
\newblock Can quantum-mechanical description of physical reality be considered
  complete?
\newblock {\em Phys. Rev.}, 47:777--780, May 1935.

\bibitem{bell_onEPR_1964}
J.~S. Bell.
\newblock On the einstein podolsky rosen paradox.
\newblock {\em Physics Physique Fizika}, 1:195--200, Nov 1964.

\bibitem{nielsen_quantum_2010}
Michael~A. Nielsen and Isaac~L. Chuang.
\newblock {\em Quantum Computation and Quantum Information}.
\newblock {Cambridge University Press}, {Cambridge ; New York}, 10th
  anniversary ed edition, 2010.

\bibitem{giovannetti_advances_2011}
Vittorio Giovannetti, Seth Lloyd, and Lorenzo Maccone.
\newblock Advances in quantum metrology.
\newblock {\em Nature Photonics}, 5(4):222--229, April 2011.

\bibitem{freedman_Experimental_1972}
Stuart~J. Freedman and John~F. Clauser.
\newblock Experimental test of local hidden-variable theories.
\newblock {\em Phys. Rev. Lett.}, 28:938--941, Apr 1972.

\bibitem{kwiat_SPDC_1995}
Paul~G. Kwiat, Klaus Mattle, Harald Weinfurter, Anton Zeilinger, Alexander~V.
  Sergienko, and Yanhua Shih.
\newblock New high-intensity source of polarization-entangled photon pairs.
\newblock {\em Phys. Rev. Lett.}, 75:4337--4341, Dec 1995.

\bibitem{hagley_Generation_1997}
E.~Hagley, X.~Ma\^{\i}tre, G.~Nogues, C.~Wunderlich, M.~Brune, J.~M. Raimond,
  and S.~Haroche.
\newblock Generation of einstein-podolsky-rosen pairs of atoms.
\newblock {\em Phys. Rev. Lett.}, 79:1--5, Jul 1997.

\bibitem{turchette_Deterministic_1998}
Q.~A. Turchette, C.~S. Wood, B.~E. King, C.~J. Myatt, D.~Leibfried, W.~M.
  Itano, C.~Monroe, and D.~J. Wineland.
\newblock Deterministic entanglement of two trapped ions.
\newblock {\em Phys. Rev. Lett.}, 81:3631--3634, Oct 1998.

\bibitem{blinov_Observation_2004}
B.~B. Blinov, D.~L. Moehring, L.-M. Duan, and C.~Monroe.
\newblock Observation of entanglement between a single trapped atom and a
  single photon.
\newblock {\em Nature}, 428:153--157, 2004.

\bibitem{togan_Quantum_2010}
Emre Togan, Yi~Chu, Alexei~S. Trifonov, Liang Jiang, Jared Maze, Lilian
  Childress, M.~V.~Gurudev Dutt, Anders~S. Sørensen, Philip~R. Hemmer,
  Alexander~S. Zibrov, and Mikhail~D. Lukin.
\newblock Quantum entanglement between an optical photon and a solid-state spin
  qubit.
\newblock {\em Nature}, 466:730--734, 2010.

\bibitem{pashkin_Quantum_2003}
Yu.~A. Pashkin, T.~Yamamoto, O.~Astafiev, Y.~Nakamura, D.~V. Averin, and J.~S.
  Tsai.
\newblock Quantum oscillations in two coupled charge qubits.
\newblock {\em Nature}, 421:823--826, 2003.

\bibitem{julsgaard_Experimental_2001}
Brian Julsgaard, Alexander Kozhekin, and Eugene~S. Polzik.
\newblock Experimental long-lived entanglement of two macroscopic objects.
\newblock {\em Nature}, 413:400--403, 2001.

\bibitem{delteil_Generation_2016}
Aymeric Delteil, Zhe Sun, Wei bo~Gao, Emre Togan, Stefan Faelt, and Ataç
  Imamoğlu.
\newblock Generation of heralded entanglement between distant hole spins.
\newblock {\em Nature Physics}, 12:218--223, 2016.

\bibitem{lee_Entangling_2011}
K.~C. Lee, M.~R. Sprague, B.~J. Sussman, J.~Nunn, N.~K. Langford, X.-M. Jin,
  T.~Champion, P.~Michelberger, K.~F. Reim, D.~England, D.~Jaksch, and I.~A.
  Walmsley.
\newblock Entangling macroscopic diamonds at room temperature.
\newblock {\em Science}, 334(6060):1253--1256, 2011.

\bibitem{lin_Quantum_2020}
Yiheng Lin, David~R. Leibrandt, Dietrich Leibfried, and Chin wen Chou.
\newblock Quantum entanglement between an atom and a molecule.
\newblock {\em Nature}, 581:273--277, 2020.

\bibitem{holland_On-demand_2023}
Connor~M. Holland, Yukai Lu, and Lawrence~W. Cheuk.
\newblock On-demand entanglement of molecules in a reconfigurable optical
  tweezer array.
\newblock {\em Science}, 382(6675):1143--1147, 2023.

\bibitem{bao_Dipolar_2023}
Yicheng Bao, Scarlett~S. Yu, Loïc Anderegg, Eunmi Chae, Wolfgang Ketterle,
  Kang-Kuen Ni, and John~M. Doyle.
\newblock Dipolar spin-exchange and entanglement between molecules in an
  optical tweezer array.
\newblock {\em Science}, 382(6675):1138--1143, 2023.

\bibitem{atlas_Observation_2024}
{The ATLAS Collaboration}.
\newblock Observation of quantum entanglement with top quarks at the atlas
  detector.
\newblock {\em Nature}, 633:542--547, 2024.

\bibitem{henke_Probing_2025}
Jan-Wilke Henke, Hao Jeng, and Claus Ropers.
\newblock Probing electron-photon entanglement using a quantum eraser.
\newblock {\em Physical Review A}, 111(1):012610, January 2025.

\bibitem{kazakevich_Spatial_2024}
Eitan Kazakevich, Hadar Aharon, and Ofer Kfir.
\newblock Spatial electron-photon entanglement.
\newblock {\em Physical Review Research}, 6(4):043033, October 2024.

\bibitem{rembold_StateAgnostic_2025a}
Phila Rembold, Santiago {Beltr{\'a}n-Romero}, Alexander Preimesberger, Sergei
  Bogdanov, Isobel~C. Bicket, Nicolai Friis, Elizabeth Agudelo, Dennis
  R{\"a}tzel, and Philipp Haslinger.
\newblock State-{{Agnostic Approach}} to {{Certifying Electron-Photon
  Entanglement}} in {{Electron Microscopy}}, February 2025.

\bibitem{dwyer_Quantum_2023}
Christian Dwyer.
\newblock Quantum {{Limits}} of {{Transmission Electron Microscopy}}.
\newblock {\em Physical Review Letters}, 130(5):056101, January 2023.

\bibitem{kruit_Designs_2016}
P.~Kruit, R.~G. Hobbs, C-S. Kim, Y.~Yang, V.~R. Manfrinato, J.~Hammer,
  S.~Thomas, P.~Weber, B.~Klopfer, C.~Kohstall, T.~Juffmann, M.~A. Kasevich,
  P.~Hommelhoff, and K.~K. Berggren.
\newblock Designs for a quantum electron microscope.
\newblock {\em Ultramicroscopy}, 164:31--45, May 2016.

\bibitem{koppell_Transmission_2022}
Stewart~A. Koppell, Yonatan Israel, Adam~J. Bowman, Brannon~B. Klopfer, and
  M.~A. Kasevich.
\newblock Transmission electron microscopy at the quantum limit.
\newblock {\em Applied Physics Letters}, 120(19):190502, May 2022.

\bibitem{compton_Quantum_1923}
Arthur~H. Compton.
\newblock A quantum theory of the scattering of x-rays by light elements.
\newblock {\em Phys. Rev.}, 21:483--502, May 1923.

\bibitem{fan_Measurement_2023}
X.~Fan, T.~G. Myers, B.~A.~D. Sukra, and G.~Gabrielse.
\newblock Measurement of the electron magnetic moment.
\newblock {\em Phys. Rev. Lett.}, 130:071801, Feb 2023.

\bibitem{barwick_photon-induced_2009}
Brett Barwick, David~J. Flannigan, and Ahmed~H. Zewail.
\newblock Photon-induced near-field electron microscopy.
\newblock {\em Nature}, 462(7275):902--906, December 2009.

\bibitem{park_Photoninduced_2010}
Sang~Tae Park, Milo Lin, and Ahmed~H Zewail.
\newblock Photon-induced near-field electron microscopy ({{PINEM}}):
  Theoretical and experimental.
\newblock {\em New Journal of Physics}, 12(12):123028, December 2010.

\bibitem{garcia_de_abajo_multiphoton_2010}
F.~Javier García~de Abajo, Ana Asenjo-Garcia, and Mathieu Kociak.
\newblock Multiphoton {Absorption} and {Emission} by {Interaction} of {Swift}
  {Electrons} with {Evanescent} {Light} {Fields}.
\newblock {\em Nano Letters}, 10(5):1859--1863, May 2010.

\bibitem{piazza_simultaneous_2015}
L~Piazza, T.T.A. Lummen, E~Qui{\~n}onez, Y~Murooka, B.W. Reed, B~Barwick, and
  F~Carbone.
\newblock Simultaneous observation of the quantization and the interference
  pattern of a plasmonic near-field.
\newblock {\em Nature Communications}, 6(1):6407, May 2015.

\bibitem{henke_integrated_2021}
Jan-Wilke Henke, Arslan~Sajid Raja, Armin Feist, Guanhao Huang, Germaine Arend,
  Yujia Yang, F.~Jasmin Kappert, Rui~Ning Wang, Marcel M{\"o}ller, Jiahe Pan,
  Junqiu Liu, Ofer Kfir, Claus Ropers, and Tobias~J. Kippenberg.
\newblock Integrated photonics enables continuous-beam electron phase
  modulation.
\newblock {\em Nature}, 600(7890):653--658, December 2021.

\bibitem{dahan_imprinting_2021}
Raphael Dahan, Alexey Gorlach, Urs Haeusler, Aviv Karnieli, Ori Eyal, Peyman
  Yousefi, Mordechai Segev, Ady Arie, Gadi Eisenstein, Peter Hommelhoff, and
  Ido Kaminer.
\newblock Imprinting the quantum statistics of photons on free electrons.
\newblock {\em Science}, 373(6561):eabj7128, August 2021.

\bibitem{gaida_Attosecond_2024}
John~H. Gaida, Hugo {Louren{\c c}o-Martins}, Murat Sivis, Thomas Rittmann,
  Armin Feist, F.~Javier {Garc{\'i}a de Abajo}, and Claus Ropers.
\newblock Attosecond electron microscopy by free-electron homodyne detection.
\newblock {\em Nature Photonics}, pages 1--7, February 2024.

\bibitem{feist_quantum_2015}
Armin Feist, Katharina~E. Echternkamp, Jakob Schauss, Sergey~V. Yalunin, Sascha
  Schäfer, and Claus Ropers.
\newblock Quantum coherent optical phase modulation in an ultrafast
  transmission electron microscope.
\newblock {\em Nature}, 521(7551):200--203, May 2015.

\bibitem{freimund_Bragg_2002}
Daniel~L. Freimund and Herman Batelaan.
\newblock Bragg {{Scattering}} of {{Free Electrons Using}} the {{Kapitza-Dirac
  Effect}}.
\newblock {\em Physical Review Letters}, 89(28):283602, December 2002.

\bibitem{priebe_Attosecond_2017a}
Katharina~E. Priebe, Christopher Rathje, Sergey~V. Yalunin, Thorsten Hohage,
  Armin Feist, Sascha Sch{\"a}fer, and Claus Ropers.
\newblock Attosecond electron pulse trains and quantum state reconstruction in
  ultrafast transmission electron microscopy.
\newblock {\em Nature Photonics}, 11(12):793--797, December 2017.

\bibitem{morimoto_Diffraction_2018}
Yuya Morimoto and Peter Baum.
\newblock Diffraction and microscopy with attosecond electron pulse trains.
\newblock {\em Nature Physics}, 14(3):252--256, March 2018.

\bibitem{vanacore_Attosecond_2018}
G.~M. Vanacore, I.~Madan, G.~Berruto, K.~Wang, E.~Pomarico, R.~J. Lamb,
  D.~McGrouther, I.~Kaminer, B.~Barwick, F.~Javier {Garc{\'i}a de Abajo}, and
  F.~Carbone.
\newblock Attosecond coherent control of free-electron wave functions using
  semi-infinite light fields.
\newblock {\em Nature Communications}, 9(1):2694, July 2018.

\bibitem{madan_Ultrafast_2022}
Ivan Madan, Veronica Leccese, Adam Mazur, Francesco Barantani, Thomas LaGrange,
  Alexey Sapozhnik, Phoebe~M. Tengdin, Simone Gargiulo, Enzo Rotunno,
  Jean-Christophe Olaya, Ido Kaminer, Vincenzo Grillo, F.~Javier~Garc{\'i}a {de
  Abajo}, Fabrizio Carbone, and Giovanni~Maria Vanacore.
\newblock Ultrafast {{Transverse Modulation}} of {{Free Electrons}} by
  {{Interaction}} with {{Shaped Optical Fields}}.
\newblock {\em ACS Photonics}, 9(10):3215--3224, October 2022.

\bibitem{fang_Structured_2024}
Yiqi Fang, Joel Kuttruff, David Nabben, and Peter Baum.
\newblock Structured electrons with chiral mass and charge.
\newblock {\em Science}, 385(6705):183--187, July 2024.

\bibitem{bendana_single-photon_2011}
Xes{\'u}s Benda{\~n}a, Albert Polman, and F.~Javier {Garc{\'i}a de Abajo}.
\newblock Single-{{Photon Generation}} by {{Electron Beams}}.
\newblock {\em Nano Letters}, 11(12):5099--5103, December 2011.

\bibitem{feist_cavity-mediated_2022}
Armin Feist, Guanhao Huang, Germaine Arend, Yujia Yang, Jan-Wilke Henke,
  Arslan~Sajid Raja, F.~Jasmin Kappert, Rui~Ning Wang, Hugo {Louren{\c
  c}o-Martins}, Zheru Qiu, Junqiu Liu, Ofer Kfir, Tobias~J. Kippenberg, and
  Claus Ropers.
\newblock Cavity-mediated electron-photon pairs.
\newblock {\em Science}, 377(6607):777--780, August 2022.

\bibitem{varkentina_cathodoluminescence_2022}
Nadezda Varkentina, Yves Auad, Steffi~Y. Woo, Alberto Zobelli, Laura Bocher,
  Jean-Denis Blazit, Xiaoyan Li, Marcel Tenc{\'e}, Kenji Watanabe, Takashi
  Taniguchi, Odile St{\'e}phan, Mathieu Kociak, and Luiz H.~G. Tizei.
\newblock Cathodoluminescence excitation spectroscopy: {{Nanoscale}} imaging of
  excitation pathways.
\newblock {\em Science Advances}, 8(40):eabq4947, October 2022.

\bibitem{arend_Electrons_2024}
Germaine Arend, Guanhao Huang, Armin Feist, Yujia Yang, Jan-Wilke Henke, Zheru
  Qiu, Hao Jeng, Arslan~Sajid Raja, Rudolf Haindl, Rui~Ning Wang, Tobias~J.
  Kippenberg, and Claus Ropers.
\newblock Electrons herald non-classical light, September 2024.

\bibitem{ben_hayun_shaping_2021}
A.~Ben~Hayun, O.~Reinhardt, J.~Nemirovsky, A.~Karnieli, N.~Rivera, and
  I.~Kaminer.
\newblock Shaping quantum photonic states using free electrons.
\newblock {\em Science Advances}, 7(11):eabe4270, March 2021.

\bibitem{dahan_creation_2023}
Raphael Dahan, Gefen Baranes, Alexey Gorlach, Ron Ruimy, Nicholas Rivera, and
  Ido Kaminer.
\newblock Creation of {Optical} {Cat} and {GKP} {States} {Using} {Shaped}
  {Free} {Electrons}.
\newblock {\em Physical Review X}, 13, July 2023.

\bibitem{zhang_Spontaneous_2024}
Bin Zhang, Reuven Ianconescu, Aharon Friedman, Jacob Scheuer, Mikhail Tokman,
  Yiming Pan, and Avraham Gover.
\newblock Spontaneous photon emission by shaped quantum electron wavepackets
  and the {{QED}} origin of bunched electron beam superradiance.
\newblock {\em Reports on Progress in Physics}, 88(1):017601, December 2024.

\bibitem{di_giulio_Tunable_2025}
Valerio~Di Giulio, Rudolf Haindl, and Claus Ropers.
\newblock Tunable quantum light by modulated free electrons, January 2025.

\bibitem{garcia_de_abajo_optical_2010}
F.~J. {Garc{\'i}a de Abajo}.
\newblock Optical excitations in electron microscopy.
\newblock {\em Reviews of Modern Physics}, 82(1):209--275, February 2010.

\bibitem{polman_electron-beam_2019}
Albert Polman, Mathieu Kociak, and F.~Javier {Garc{\'i}a de Abajo}.
\newblock Electron-beam spectroscopy for nanophotonics.
\newblock {\em Nature Materials}, 18(11):1158--1171, November 2019.

\bibitem{ma_Delayedchoice_2016}
Xiao-song Ma, Johannes Kofler, and Anton Zeilinger.
\newblock Delayed-choice gedanken experiments and their realizations.
\newblock {\em Reviews of Modern Physics}, 88(1):015005, March 2016.

\bibitem{scully_Quantum_1982}
Marlan~O. Scully and Kai Dr\"uhl.
\newblock Quantum eraser: A proposed photon correlation experiment concerning
  observation and "delayed choice" in quantum mechanics.
\newblock {\em Phys. Rev. A}, 25:2208--2213, Apr 1982.

\bibitem{harvey_efficient_2014}
Tyler R~Harvey, Jordan S~Pierce, Amit K~Agrawal, Peter Ercius, Martin Linck,
  and Benjamin~J McMorran.
\newblock Efficient diffractive phase optics for electrons.
\newblock {\em New Journal of Physics}, 16(9):093039, September 2014.

\bibitem{johnson_scanning_2021}
Cameron~W. Johnson, Amy~E. Turner, and Benjamin~J. McMorran.
\newblock Scanning two-grating free electron {{Mach-Zehnder}} interferometer.
\newblock {\em Physical Review Research}, 3(4):043009, October 2021.

\bibitem{taleb_Phaselocked_2023}
Masoud Taleb, Mario Hentschel, Kai Rossnagel, Harald Giessen, and Nahid Talebi.
\newblock Phase-locked photon--electron interaction without a laser.
\newblock {\em Nature Physics}, 19(6):869--876, June 2023.

\bibitem{lvovsky_Iterative_2004}
A~I Lvovsky.
\newblock Iterative maximum-likelihood reconstruction in quantum homodyne
  tomography.
\newblock {\em Journal of Optics B: Quantum and Semiclassical Optics},
  6(6):S556, may 2004.

\bibitem{kfir_entanglements_2019}
Ofer Kfir.
\newblock Entanglements of {{Electrons}} and {{Cavity Photons}} in the
  {{Strong-Coupling Regime}}.
\newblock {\em Physical Review Letters}, 123(10):103602, September 2019.

\bibitem{konecna_entangling_2022}
Andrea Kone{\v c}n{\'a}, Fadil Iyikanat, and F.~Javier {Garc{\'i}a de Abajo}.
\newblock Entangling free electrons and optical excitations.
\newblock {\em Science Advances}, 8(47):eabo7853, November 2022.

\bibitem{baranes_free_2022}
Gefen Baranes, Ron Ruimy, Alexey Gorlach, and Ido Kaminer.
\newblock Free electrons can induce entanglement between photons.
\newblock {\em npj Quantum Information}, 8(1):1--9, March 2022.

\bibitem{haindl_Coulomb_2023}
Rudolf Haindl, Armin Feist, Till Domr\"{o}se, Marcel M\"{o}ller, John~H. Gaida,
  Sergey~V. Yalunin, and Claus Ropers.
\newblock Coulomb-correlated electron number states in a transmission electron
  microscope beam.
\newblock {\em Nature Physics}, 19:1410--1417, June 2023.

\bibitem{lichte_Inelastic_2000}
Hannes Lichte and Bert Freitag.
\newblock Inelastic electron holography.
\newblock {\em Ultramicroscopy}, 81(3):177--186, 2000.

\bibitem{potapov_Inelastic_2007}
P.L. Potapov, J.~Verbeeck, P.~Schattschneider, H.~Lichte, and D.~Van~Dyck.
\newblock Inelastic electron holography as a variant of the {{Feynman}} thought
  experiment.
\newblock {\em Ultramicroscopy}, 107(8):559--567, August 2007.

\bibitem{mechel_Quantum_2021}
Chen Mechel, Yaniv Kurman, Aviv Karnieli, Nicholas Rivera, Ady Arie, and Ido
  Kaminer.
\newblock Quantum correlations in electron microscopy.
\newblock {\em Optica}, 8(1):70--78, January 2021.

\bibitem{lu_Bayesian_2022}
Hsuan-Hao Lu, Karthik~V. Myilswamy, Ryan~S. Bennink, Suparna Seshadri,
  Mohammed~S. Alshaykh, Junqiu Liu, Tobias~J. Kippenberg, Daniel~E. Leaird,
  Andrew~M. Weiner, and Joseph~M. Lukens.
\newblock Bayesian tomography of high-dimensional on-chip biphoton frequency
  combs with randomized measurements.
\newblock {\em Nature Communications}, 13:4338, 2022.

\bibitem{kim_Bayesian_2023}
Ki-Tae Kim, Umberto Villa, Matthew Parno, Youssef Marzouk, Omar Ghattas, and
  Noemi Petra.
\newblock hippylib-muq: A bayesian inference software framework for integration
  of data with complex predictive models under uncertainty.
\newblock {\em ACM Trans. Math. Softw.}, 49(2), June 2023.

\end{thebibliography}


\clearpage
\onecolumngrid

\appendix

\renewcommand{\thesection}{SI\arabic{section}}
\renewcommand{\thefigure}{SI\arabic{figure}}
\setcounter{figure}{0}

\section*{Supplementary Information}

\section{Experimental setup}

\begin{figure*}[ht]
    \centering
    \includegraphics[width=\textwidth]{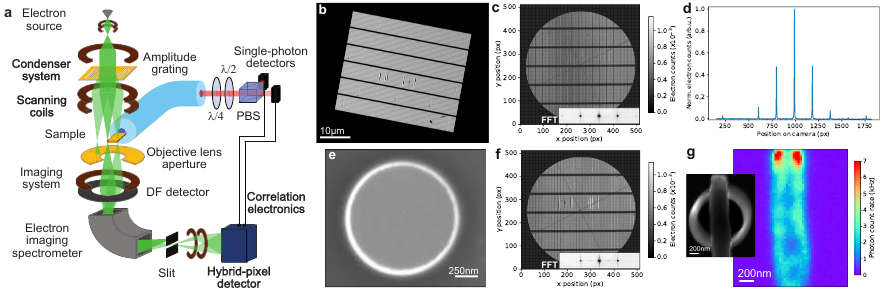}
    \caption{\textbf{Experimental setup}
        \textbf{a} Extended schematic drawing of the experimental setup: The electron beam (green) is provided by a cold field-emitter and coherently by an amplitude grating placed in the condenser lens system.
        Employing the scanning coils, the electron beams are positioned such that two of them pass by a sample structure (gold) placed in front of an optical single-mode fiber (blue) and can generate photons (red).
        All further electron beams not interacting with the specimen are blocked by an additional aperture placed inside the objective lens right below the sample.
        The generated photons are collected via the optical fiber and guided to the polarisation state-sensitive detection setup.
        In parallel, the electron beams are recombined by the imaging lenses and passed through an imaging spectrometer (grey) that allows for an energy filtering of electrons using an energy-selective slit in the dispersive plane of the spectrometer, while scattered electrons are recorded on a dark-field detector (DF).
        Electrons that have lost energy corresponding to the generation of a photon are then detected on a hybrid-pixel camera (dark blue), yielding both timing and position information.
        Linking these two detection setups enables electron-photon correlation measurements.
        \textbf{b} Scanning electron microscope image of the amplitude grating prepared by FIB milling of a gold-coated silicon nitride membrane and employed as an electron beam splitter.
        \textbf{c} Elastic interference pattern as recorded on the hybrid-pixel detector placed behind the imaging spectrometer without energy filtering and with the electron beams placed far away from the sample. The inset shows the Fourier transform (FFT) of the pattern, exhibiting five visible spots and, thus, indicating the contribution of more than two partial beams to the electron pattern.
        \textbf{d} Electron probes formed by the amplitude grating in the sample plane and imaged on a camera in front of the imaging spectrometer. The relative intensity in the 0th and 1st order beam is found to be \unit[64]{\%} and \unit[36]{\%}, respectively.
        \textbf{e} Scanning electron microscope of the custom objective lens aperture placed below the sample to block the additional electron beams produced by the amplitude mask.
        \textbf{f} Elastic interference pattern recorded with the post-spectrometer detector in the absence of the energy-selective slit and with the electron beams positioned away from the sample. The Fourier transform (FFT, inset) exhibits only three peaks, indicating that only two beams pass the additional aperture to form the interference pattern.
        \textbf{g} Map of the photon count rate obtained while scanning a single electron beam across the sample and the objective lens aperture as shown in the dark-field image recorded in parallel. No increased count rate is observed when the electron beam hits the aperture.
    }
    \label{fig:SI1}
\end{figure*}

This section aims to provide additional information and data on the experimental setup employed in the study.
Our experimental realisation, outlined in Figure \ref{fig:1} of the main text and illustrated in more detail in Figure \ref{fig:SI1}a, is based on a conventional transmission electron microscope (TEM) equipped with a cold-field emission electron source operated at \unit[80]{keV} mean electron energy and a spread of approximately \unit[500]{meV}.

The TEM is modified to allow for coherent splitting of the electron beam by means of an amplitude grating placed in the condenser lens system (cf. Fig. \ref{fig:SI1}a).
This grating, fabricated from a gold-coated silicon nitride membrane via focused ion-beam milling and shown in the scanning electron microscope image in Figure \ref{fig:SI1}b.
It gives rise to multiple focused electron probes of $<\unit[20]{nm}$ diameter spaced by approx. \unit[510]{nm} in the sample plane of the microscope (cf. Fig. \ref{fig:SI1}d) as imaged on a camera below the imaging lenses.
From this image, the ratio of intensities in the 0th and 1st order beams passing the sample is determined to be \unit[64]{\%} and \unit[36]{\%}, respectively.
For our measurements, the beams are recombined using the imaging lenses and passed through an imaging spectrometer that allows for selecting electrons that have lost energy corresponding to that of a photon.
The resulting interference pattern is imaged on a hybrid-pixel detector and Figure \ref{fig:SI1}c shows the elastic interference pattern (i.e. without the energy-selective slit in the dispersive plane of the spectrometer) with the beams positioned far off the sample.
As apparent from the additional second-order peaks in the FFT of this interference pattern (cf. inset in Fig. \ref{fig:SI1}c), more than two electron beams are contributing.

The grating-generated electron beams that do not interact with the sample are blocked by a custom $\unit[1.2]{\mu m}$ diameter aperture placed in the objective lens pole piece right below the sample (cf. Fig. \ref{fig:SI1}a).
This has direct impact on the elastic interference pattern, recorded without the energy-selective slit of the spectrometer inserted and with the electron beams far off the sample (cf. Fig. \ref{fig:SI1}f), whose Fourier transform only consists of three spots in accordance to the interference of two beams.
From this data, the coherence of the input electron beam is estimated to be \textcolor{red}{???} considering the interference fringe visibility as well as the intensity ratios of the 0th and 1st order beams. 
To ensure that electrons hitting the aperture do not generate photons that are collected via the optical fiber and could lead to false electron-photon coincidences, a single electron probe is scanned across both the sample and the aperture as apparent from the dark-field image recorded in parallel (cf. inset in Fig. \ref{fig:SI1}g).
The resulting map of the photon count rate, presented in Fig. \ref{fig:SI1}g, does not exhibit increased counts outside the sample region.

\section{Sample characterisation}

This section aims to provide extended information on the emission spectrum of the electron-driven photon generation.

\begin{figure*}[ht]
   \centering
   \includegraphics[width=0.75\textwidth]{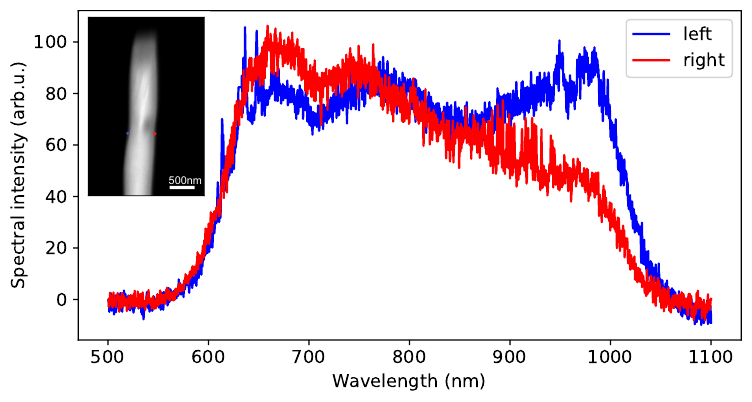}
   \caption{\textbf{Cathodoluminescence spectra}
       Optical spectra of electron-driven photon generation at an electron energy of $\unit[200]{keV}$ as recorded with a grating spectrometer.
       The two spectra are collected with the beam, formed by a $\unit[200]{\mu m}$ condenser aperture, on the left (blue) and right (red) side of the metal-coated glass prism, indicated by the coloured spots in the inset dark-field image.
   }
   \label{fig:SI2}
\end{figure*}

Extended characterisation of the sample structure was performed with a single electron beam at \unit[200]{keV} center energy and using a larger $\unit[200]{\mu m}$ condenser aperture to increase the electron current.
Optical spectra were recorded with the output fiber directly connected to an Acton Research Corporation SpectraPro-2500i grating spectrometer fitted with a liquid-nitrogen cooled CCD camera (Princeton Instruments Spec-10:100) and using a grating with \unit[150]{lines/mm} and an integration time of \unit[30]{s} per spectrum.
The spectra obtained for the beam on the left (blue) or right (red) side of the sample (cf. inset in Fig. \ref{fig:SI2}) are shown in Figure \ref{fig:SI2}, with the decrease in intensity at large wavelengths linked to the decrease in the grating and detector efficiencies.
Overall, the emission behaviour of photons generated on the two sides is similar with the differences mostly linked to differences in the grating efficiency for s- and p-polarised light.

\section{Data processing \& evaluation}

This section aims to provide additional information and figures on the data processing, covering the coincidence detection and background correction, as well as the data evaluation including the Bayesian reconstruction and the coherence correction.

\begin{figure*}[ht]
   \centering
   \includegraphics[width=0.95\textwidth]{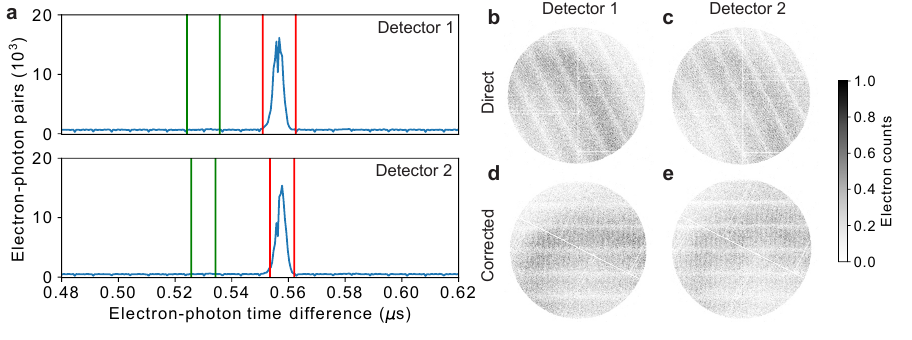}
   \caption{\textbf{Coincidence filtering, background subtraction and rotation}
   \textbf{a} Electron-photon coincidences are observed as a peak in the electron-photon time difference histogram for each of the photon detectors with the delay determined by the experimental setup. The employed coincidence window (red) allows for selecting electrons that generated a photon detected on one of the detectors. The background correction data is obtained from a number of time windows (one illustrated in green) outside this coincidence time frame.
   \textbf{b\&c} Energy- and coincidence-filtered electron interference patterns without background correction for coincidences with detectors 1 and 2, respectively.
   \textbf{d\&e} Performing the background correction by subtracting the average counts outside the coincidence window and rotating the image to align the interference fringes vertically results in the displayed patterns for coincidences with detectors 1 and 2, respectively.
   }
   \label{fig:SI3}
\end{figure*}

In the interference measurements presented in Figure 3 of the main text, both photon and energy-filtered electron arrival times are registered on the TimePix detectors, harnessing the device's time-to-digital converters.
Comparing the arrival time of electrons to the detection of photons on detectors 1 and 2 results in the histograms presented in Fig. \ref{fig:SI3}a.
They exhibit a pronounced coincidence peak with the exact delay determined by the path lengths in the experimental setup.
Selecting the electrons that are in coincidence with the detection of a photon on one of the detectors, defined by the coincidence time window shown in red, and binning them spatially yields the two interference patterns shown in Fig. \ref{fig:SI3}b\&c.
These patterns are corrected for the uncorrelated background by averaging over 10 time windows of the width of the coincidence peak, one of them illustrated in green in Fig. \ref{fig:SI3}a, and subtracting the resulting distribution from the coincidence pattern.
A rotation of the images is then applied to align the interference fringes vertically, leading to the corrected, processed interference patterns presented in Fig. \ref{fig:SI3}d\&e.
The electron counts in these patterns are then grouped by their phase in the oscillation, resulting in the distribution of detection probabilities shown in the main text Fig. 3b-d.

\begin{figure*}[ht]
   \centering
   \includegraphics[width=0.95\textwidth]{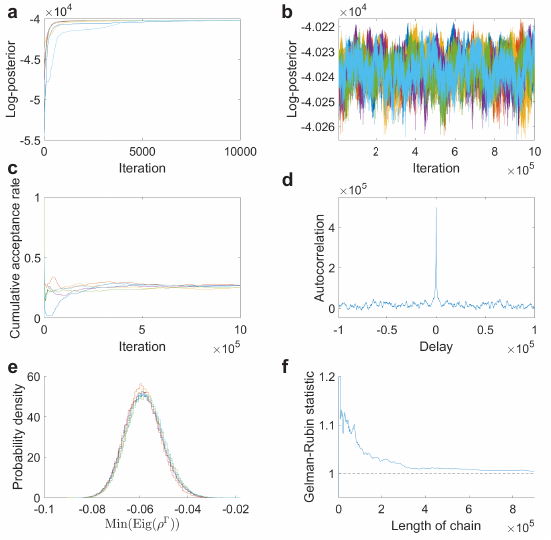}
   \caption{\textbf{Convergence of the Bayesian reconstruction.}
   \textbf{a} Log posterior of all chains from iteration 1 to 10000.
   \textbf{b} log-posterior from 1e4 to 1e6 (end).
   \textbf{c} Cumulative acceptance rate of each chain, with an average of about $26\%$.
   \textbf{d} Autocorrelation of the minimal eigenvalue of the partially transposed density matrix. A section of length 5e5 is taken from the middle of the chain and its normalised cross-correlation with chains delayed between -1e5 and 1e5 iterations is computed. The autocorrelation decays to zero after about 5000 samples.
   \textbf{e} Posterior probability density of the minimal eigenvalue of the partially transposed density matrix for each chain.
   \textbf{f} Evolution of the Gelman-Rubin test statistic according to the length of the chains. All 6 chains are used in the calculation of the statistic, with $10\%$ burn-in. 
   }
   \label{fig:SI4}
\end{figure*}

Figure \ref{fig:SI4} contains the extended data on the convergence of the Bayesian reconstruction of the electron-photon quantum state discussed in the Methods section of the main text.

\begin{figure*}[ht]
   \centering
   \includegraphics[width=\textwidth]{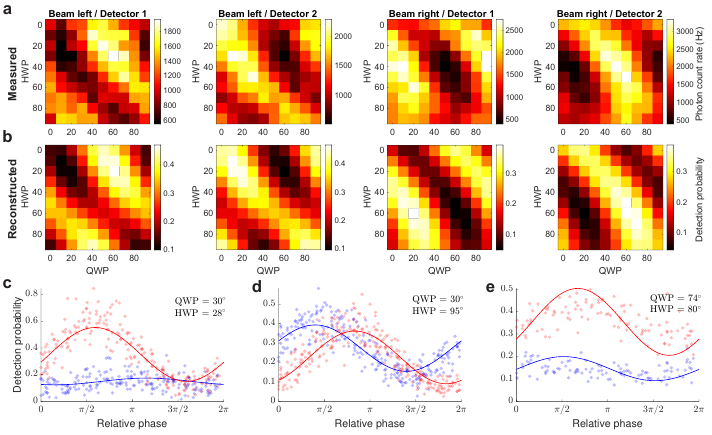}
   \caption{\textbf{Consistency of posterior means with observed data.} 
   The Bayesian reconstructed density matrix represents the observations very well. \textbf{a,b} The observed CL counts for various waveplate settings, and when the electron beam illuminates either side of the sample (a) agree with the Bayesian posterior mean for the probabilities (b) at the same settings.
   \textbf{c-e} The posterior mean (solid lines) also describes the observed electron interference patterns (data points).
   }
   \label{fig:SI5}
\end{figure*}

Figure \ref{fig:SI5} presents a comparison of the results of the Bayesian reconstruction with the measured photon count-rates (cf. Fig. \ref{fig:SI5}a\&b) as well as with the electron interference patterns recorded at different wave plate settings (cf. Fig. \ref{fig:SI5}c-e).

\begin{figure*}[ht]
   \centering
   \includegraphics[width=0.6\textwidth]{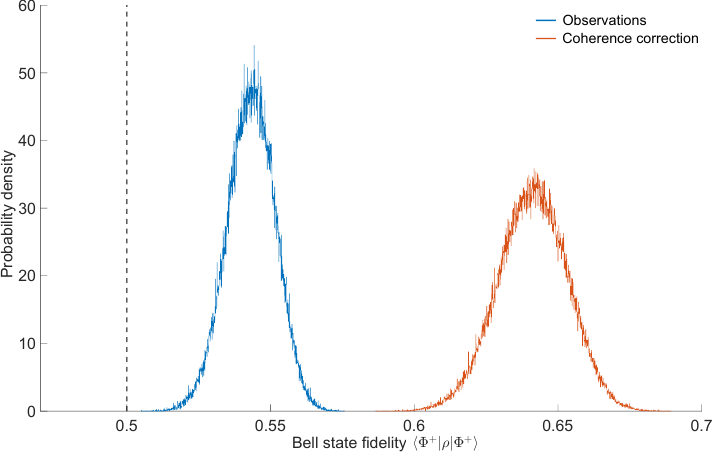}
   \caption{\textbf{Corrections for non-ideal spatial coherence of the initial electron state.} The initial coherence is determined from the electron interference patterns off the sample (cf. Fig. \ref{fig:SI1}f) to be $\gamma = 72.7\%$. Correcting for this results in an increased fidelity with respect to the Bell state $\ket{\Phi^+}$ of $0.64$ compared to the uncorrected value of $0.54$.
   }
   \label{fig:SI6}
\end{figure*}

Figure \ref{fig:SI6} shows the impact of the coherence correction, discussed in the main text, on the fidelity of the reconstructed electron-photon state with respect to the Bell state $\ket{\Phi^+}$, increasing the mean value from \unit[0.54]{} to \unit[0.64]{}.

\end{document}